\documentclass[conference]{IEEEtran}
\usepackage{graphicx}
\usepackage{diagbox}
\usepackage{cite}
\usepackage{amsmath,amssymb,amsfonts}
\usepackage{algorithm}
\usepackage{algorithmic}

\usepackage{graphicx}
\usepackage{textcomp}
\usepackage{xcolor}
\usepackage{url}
\usepackage{verbatim}
\usepackage{multirow}
\usepackage{soul}

\usepackage{listings}

\ifCLASSOPTIONcompsoc
    \usepackage[caption=false, font=normalsize, labelfont=sf, textfont=sf]{subfig}
\else
\usepackage[caption=false, font=footnotesize]{subfig}
\fi

\usepackage{lipsum,afterpage,refcount}

\begin{document}
\title{Extending Desbordante with Probabilistic Functional Dependency Discovery Support}
\date{}

\author{
\IEEEauthorblockN{Ilia Barutkin, Maxim Fofanov, Sergey Belokonny, Vladislav Makeev, George Chernishev}
\IEEEauthorblockA{Saint-Petersburg University \\ Saint-Petersburg, Russia \\ \{ilia.d.barutkin, max.fofanov, belokoniy,  makeev.vladislav.d, chernishev\}@gmail.com\\}}


\maketitle

\begin{abstract}

Data profiling aims to extract complex patterns from data for further analysis and use that data in domains such as data cleaning, data deduplication, anomaly detection, and many more.

Functional dependencies (FDs) are one of the most well-known patterns. However, they are poorly suited for these tasks, as real data is usually dirty, and the rigid definition of FDs does not allow algorithms to locate them. For this reason, there are several formulations aimed at relaxing FDs to support dirty data, with approximate functional dependency (AFD) being the most popular one. Another formulation is the Probabilistic Functional Dependency (pFD), which we aim to support inside Desbordante~--- a science-intensive, high-performance and open-source data profiling tool implemented in C++. However, pFDs are relatively poorly studied, compared to AFDs.

In this paper we study pFDs, both analytically and empirically. We start by assessing how different pFDs and AFDs are by studying cases in which pFDs have an edge over AFDs. Then, we implement the algorithm for pFD discovery, as well as study its run time and memory consumption. We also compare it with an AFD discovery algorithm. Lastly, we study the output of both algorithms to learn whether or not it is possible to use AFD discovery algorithm to get pFDs and vice versa.
\end{abstract}

\section{Introduction}

Currently, growing volumes of data pose a serious challenge to data analysts. Data, however, offers only a moderate value in and of itself, and it is instead the facts contained within that data that are of interest to analysts. The volumes of data in question far exceed the size that could be grasped by the human eye, so automatic approaches become more and more in demand. 

Data profiling~\cite{10.5555/3312004} aims to extract facts from data. There are two kinds of data profiling~--- naive and science-intensive. Naive approach concerns itself with simple statistics, such as the number of rows and columns, number of nulls in them, their mean and variance, etc. There are dozens of tools for this kind of profiling. On the other hand, science-intensive profiling aims to extract complex patterns represented by structures which we will refer to as primitives. Examples of such patterns are database dependencies (functional~\cite{10.14778/2794367.2794377}, inclusion~\cite{10.1145/3357384.3357916}), association rules~\cite{10.5555/2677098}, algebraic constraints~\cite{10.5555/1315451.1315509}, inferred semantic data types~\cite{10.1145/3292500.3330993}, and others. Such patterns have many applications:
\begin{itemize}
    \item for scientific data, they may indicate a presence of some regularity~\cite{10.1145/3459930.3469499}, which may promote the formulation of a hypothesis, which, in turn, may lead to a scientific discovery;
    \item for business data, it is possible~\cite{10.14778/3377369.3377379} to use the discovered primitives for cleaning errors in data, finding inexact duplicates, performing schema matching, finding outliers, and solving many other problems;
    \item for machine learning, data primitives can help in feature engineering and in choosing the direction for the ablation study;
    \item for databases, they can help with validating and discovering various advanced integrity constraints.
\end{itemize}

Extracting and validating primitives is computationally expensive, which becomes a serious issue with the scaling of datasets. Therefore, it requires complex algorithms and efficient implementations. These are some of the major contributing factors as to why such kind of profiling is now a developing area and why science-intensive profilers are rare. Currently, there exist two science-intensive data profilers~--- Metanome~\cite{10.14778/2824032.2824086} and Desbordante.

Desbordante (Spanish for \textit{boundless})~\cite{DBLP:journals/corr/abs-2301-05965} is a \textit{science-intensive}, \textit{high-performance} and \textit{open-source} data profiling tool implemented in C++. To the best of our knowledge, Desbordante is currently the only profiler that possesses these three qualities. It is capable of discovering and validating many primitives, including functional dependencies (both exact and approximate), conditional functional dependencies, metric functional dependencies, and others. The full list can be found on the web-site~\cite{desbordante_repo}. 

One of the well-known primitives is the functional dependency, which states that if two records of the table are equal in attribute X, then they should be equal in attribute Y. The formal definition is given in Section~\ref{sec:background}.

Primitives can be classified into three groups:
\begin{enumerate}
    \item Exact by definition. These primitives define instances which hold over the whole dataset. Classic functional dependency is an example of the exact primitive.
    \item Approximate by definition. In this case, approximate means that found instances hold over the whole dataset, but with some degree of error predefined by user at the start of the algorithm. Thus, there are records in the dataset that may not conform to the exact definition.
    \item Approximate by discovery procedure. In this case approximate means that discovery algorithm returns primitive instances that may hold or may not hold. While such instances require verification, such approach may be of use as it allows the speed up of discovery by up to an order of magnitude~\cite{10.1145/2983323.2983781, KPDFHZZN17FastApproximate}.
\end{enumerate}

In this paper, we will only consider dependencies that are approximate by definition, also called relaxed dependencies~\cite{7219433}. Such dependencies are of a particular interest for the end-users of science-intensive profilers. There is a simple reason for this: real-life data is always dirty~--- it contains inconsistencies, missing values, and other artifacts. Therefore, exact dependencies rarely hold on such data and discovery algorithm will not locate them.

For functional dependencies, there are several approximate variants that are built upon the family of $g_1, g_2, g_3$ metrics, proposed by J.Kivinen and H.Mannila in their seminal paper ``Approximate inference of functional dependencies from relations''~\cite{KIVINEN1995129}. The most well-known variant is Approximate Functional Dependency (AFD)~\cite{10.14778/3192965.3192968}, which is based on an adaptation of the $g_1$ metric for defining maximum permissible error. One of the alternatives is the Probabilistic Functional Dependency~\cite{wang, wang_report}, which uses $g_3$. 

We are considering the addition of pFD discovery functionality to Desbordante. Before this, it is necessary to evaluate pFDs, since they are significantly less studied that AFDs. At the same time, Desbordante supports discovery and validation of FDs and AFDs, so it is natural to compare them with each other. Our goal is twofold: firstly, it is essential to study how expensive in terms of run time and memory consumption pFDs are when compared to exact approaches. Secondly, it is also necessary to understand if pFD support provides value to the end-user. This includes answering questions ``how different are pFDs from AFDs'', and ``how dependencies of both types that are returned by discovery algorithm relate to each other''.

Overall, in our study we pose the following research questions (RQs):

\begin{itemize}
    \item[RQ1] Are pFDs of interest to the end-user? How different are they from AFDs, what kind of FD violations are they more tolerant to? Does this definition allow for discovery of dependencies that could not be discovered by the AFD definition? And vice versa: how much AFDs are lost by it?
    \item[RQ2] How computationally expensive is the candidate validation procedure of pFD discovery algorithm compared to the AFD?
    \item[RQ3] How does maximum error threshold affect run time and memory consumption of the pFD discovery algorithm?
    \item[RQ4] What is the run time and memory expenses of pFD discovery, compared to AFD?
\end{itemize}

Overall, the contribution of the paper is the following:

\begin{itemize}
    \item A discussion of pFDs, their comparison with AFDs.
    \item A survey of approximate primitives that are based on $g_1, g_2, g_3$ metrics, as it is the basis for the majority of existing approximate primitives, including AFDs and pFDs.
    \item An open-source C++ implementation of a pFD discovery algorithm, which~--- to the best of our knowledge~--- is the only one currently available.
    \item An empirical evaluation of the pFD discovery algorithm, and its comparison with the AFD one.
\end{itemize}

This paper is organized as follows. In Section~\ref{sec:background} we formally present pFDs and AFDs. We compare them and discuss their differences, while providing examples. Next, in Section~\ref{sec:relwork} we discuss related work concerning approximate dependencies based on $g_1, g_2, g_3$ metrics. In Section~\ref{sec:algo} we describe the algorithm and discuss our modifications. We evaluate our implementation and compare pFD and AFD discovery in Section~\ref{sec:experiments}. We conclude this paper with Section~\ref{sec:concl}.

\section{Background}\label{sec:background}

Let us start with basic definitions imperative to understanding the paper's context.

A \ul{functional dependency}~\cite{tane} over a relation $r$ with schema $R$ is an expression denoted as $X \to A$, where $X \subseteq R$ and $A \in R$. We also denote set X as left-hand side (LHS), and the attribute A as right-hand side (RHS). The dependency is \ul{satisfied} if, for all pairs of tuples $t,u \in r$, the following holds: if $\forall B \in X (t[B]=u[B])$, then $t[A]=u[A]$, or, equivalently, $t$ and $u$ \ul{agree} on $X$ and $A$. In this case, we also say that the functional dependency is \ul{correct} or \ul{holds}.

Let a certain relation $r$ with schema $R$ and a functional dependency $X \to Y$ over $R$ be given. Then we assert that a pair $(u,v)$ of tuples from $r$ \ul{violates the dependency}, or equivalently, is a \ul{violating pair}, if $u[X]=v[X]$, but $u[Y] \neq v[Y]$. From this, it can be concluded that the dependency holds on the relation if the relation contains no violating pairs. A tuple $u$ is termed \ul{violating} if it is a part of a violating pair.

A \ul{relaxed functional dependency} is a functional dependency that is almost satisfied. An example of this could be the relationship between columns ``phone number'' and ``department'', as several departments within company may share the same phone, albeit rarely. There are several ways to define the relaxation of functional dependency. The first one is the notion of \ul{Approximate Functional Dependency}. In the original TANE paper~\cite{tane} authors proposed to use the $g_3$ metric to define and discover AFDs, which is as follows:
\begin{equation*}
g_{3}(X\xrightarrow{}Y, r) = 1 - \frac{max\{ |s| | s \subseteq r, s \models X \xrightarrow{}Y\}}{|r|}
\end{equation*}

Almost two decades later S. Kruse and F. Naumann~\cite{10.14778/3192965.3192968} developed PYRO~--- a novel AFD discovery algorithm. However, they have used a modified $g_1$ metric for their AFD definition, which is as follows:

\begin{equation*}\label{eq:e}
\begin{split}
e(X\xrightarrow{}&Y, r) =  \\
&\frac {|\{(t_1, t_2) \in r^2 \mid t_1[X]=t_2[X] \wedge t_1[Y] \not= t_2[Y] \}|}{|r|^2 - |r|}
\end{split}
\end{equation*}

Thus, currently there exist two algorithms for AFD discovery and two AFD definitions. Metrics which are used in these definitions can be put into both existing algorithms.

Desbordante has both TANE and PYRO implementations~\cite{9435469}. Having started our project, we have decided to stick to the modern definition, and thus in this paper we consider AFDs that are based on the modified $g_1$ metric. It is also worth mentioning that our implementation of TANE is a modified one, similarly to TANE implementation in the Metanome project~\cite{tanex_repo}.

The second relaxation approach is the \ul{Probabilistic Functional Dependency}, which is defined as follows. Let R~--- a relation, X~--- a set of attributes, and A~--- an attribute in R. A probabilistic functional dependency~\cite{wang} is denoted as $pFD: X\xrightarrow{p}Y$, where p is the likelihood of the $X\xrightarrow{} Y$ being correct.

To define said \ul{probability}, lets denote a set of unique not-null values of attributes of X as $D_{X} =\ \{t[X]  \mid t \in R\}$, set of tuples with those values of $X_1$ attributes of X as $V_{X_1}  =\ \{t \in R \mid t[X] = X_1\}$, and another set of tuples as 
$(V_Y, V_{X_1}) =\ \{t \in R \mid t[X] = X_1 \wedge t[Y] = \underset{\substack{Y_k \in D_{Y}}}{arg max} \{ | V_{X_1}\cap V_{Y_k} | \} \}$. The probability of a dependency holding on a subset of tuples with value of attribute $X$ equal to $X_1$ is therefore defined as $P(X\xrightarrow{} Y, V_{X_1}) = \frac{|V_Y, V_{X_1}|}{|V_{X_1}|}$.

Finally, the probability of a functional dependency between attributes X and Y in R is defined via two formulas, namely \ul{PerValue} and \ul{PerTuple}:

\begin{equation*}
P_{PerValue}(X\xrightarrow{} Y, R) = \frac {\sum_{V_X \in D_X} P(X\xrightarrow{} Y, V_X)} {|D_X|}
\end{equation*}
\begin{equation*}
P_{PerTuple}(X\xrightarrow{} Y, R) = \frac{\sum_{V_X \in D_X} |V_Y, V_{X}|}{\sum_{V_X \in D_X} |V_{X}|}
\end{equation*}

It is evident PerValue is an average of the probabilities of a dependency being correct for each distinct value of X, whereas PerTuple metric accounts for the frequency of values of X amongst all tuples in a relation. 

We also say that a pFD $\overline{X} \xrightarrow{p} Y$ is \ul{minimal} if, for any proper subset $X'\subset \overline{X}$, $X' \xrightarrow{p} Y$ does not hold. A pFD is called \ul{trivial}, if $Y \in \overline{X}$.

Note that it is possible to add an attribute to LHS of a pFD, and the resulting dependency will remain a pFD, if PerTuple metric is used. The same holds true for AFDs and their $g_1$ metric. However, this is not always true in case of pFD PerValue.

Finally, it is evident that PerTuple metric is the same as $g_3$, which is defined using notion of probability: 
\begin{equation*}
P_{PerTuple}(X\xrightarrow{} Y, R) = 1 - g_3(X\xrightarrow{} Y, R)
\end{equation*}

\section{Related Work}\label{sec:relwork}

In the world of relaxed dependencies, there are three major metrics used for defining how well a given relaxed dependency holds on a particular dataset. They are called $g_1, g_2, g_3$ and were proposed by J.Kivinen and H.Mannila in ``Approximate inference of functional dependencies from relations''~\cite{KIVINEN1995129}. Despite the fact that the original paper considers relaxed functional dependencies, the concept is easily generalized owing to the flexibility of the provided definitions. As the result, these metrics gave rise to many other types of relaxed dependencies which we are going to survey in this paper.

\subsection{$g_1$ and  $g_2$ metrics}
Let $G_1$ be defined as the number of violating pairs for the dependency $X \to Y$ in the relation $r$:
\begin{equation*}
\begin{split}
G_1(X \to Y, r) = |\{(u, v) | u,v \in r, & \\ 
u[X] &= v[X] \land u[Y] \neq v[Y]\}|.
\end{split}
\end{equation*}

Then, the metric $g_1$ represents a normalized version of $G_1$.
\begin{equation*}
g_1(X \to Y, r) = G_1(X \to Y, r) / |r|^2
\end{equation*}

$G_2$ is the number of violating tuples for the dependency $X \to Y$ in the relation $r$.

\begin{equation*}
\begin{split}
G_2(X \to Y, r) = |\{u | u \in r,& \\ 
\exists v \in r: &u[X] = v[X] \land u[Y] \neq v[Y]\}|
\end{split}
\end{equation*}

The metric $g_2$, in turn, represents a normalized version of $G_2$.
\begin{equation*}
g_2(X \to Y, r) = G_2(X \to Y, r) / |r|
\end{equation*}
The $g_1$ and $g_2$ metrics, as shown in the previously mentioned paper~\cite{KIVINEN1995129}, are applied for defining approximate functional dependencies. However, due to their poorer generalizability and greater computational complexity, they are not as widely used as $g_3$~\cite{vilmin2023functional}.

\begin{table}[h!]
\centering
\caption{Example of $g_1, g_2$ and $g_3$}
\label{tbl:ExampleCalculation}

\begin{center}
\begin{tabular}{|c | c|} 
\hline
X & Y \\ [0.5ex] 
\hline\hline
a & 1 \\ 
\hline
b & 2 \\ 
\hline
a & 3 \\ 
\hline
c & 3 \\ 
\hline
d & 4 \\ 
\hline
\end{tabular}
\end{center}
\end{table}

Consider an example presented in Table~\ref{tbl:ExampleCalculation}. In case of $g_1$ its value is calculated as follows. Since there is only a single violating tuple~--- $((a, 1), (a, 3))$, we get:
$$g_1(X \to Y, r) = \frac{1}{5^2} = 0.04.$$
For $g_2$, there are two values with different right-hand sides: $(a,1)$ and $(a,3)$, hence the value of the metric $g_2$ being:
$$g_2(X \to Y, r) = \frac{2}{5} = 0.4.$$

Now, let us consider various relaxed dependencies that are based on either $g_1$ or $g_2$.

\begin{table}[h!]
\caption{AFD Example: Position $\to$ Salary ($g_3 = 0.16$).}
\label{tbl:PositionToSalary}

\begin{center}
\begin{tabular}{|c | c | c|}
\hline
ID & Position & Salary \\ [0.5ex] 
\hline\hline
1 & Programmer & 3000 \\ 
\hline
2 & Designer & 2800 \\ 
\hline
3 & Programmer & 3200 \\ 
\hline
4 & Manager & 3000 \\ 
\hline
5 & Designer & 2800 \\ 
\hline
6 & Programmer & 3000 \\ 
\hline
\end{tabular}
\end{center}
\end{table}
\textbf{1. Approximate functional dependencies.} We have discussed the notion of AFDs in the Background section. An AFD example is presented in Table~\ref{tbl:PositionToSalary}.

PYRO~\cite{10.14778/3192965.3192968} is an algorithm for discovery of AFDs that are based on the modern definition. In this algorithm, an adaptation of the $g_1$ metric, referred to by the authors of the article as $e$, is employed.
\begin{align*}
e(X \to &A, r) = \\
& = \frac{|\{(t_1, t_2) \in r^2 | t_1[X] = t_2[X] \land t_1[A] \neq t_2[A]\}|}{|r|^2 - |r|}.
\end{align*}

PYRO demonstrates excellent performance due to employing several interesting optimizations, one of them being the error calculation approach.

Let $r$ be a relation with schema $R$ and $X \subseteq R$ be a set of attributes. A \ul{cluster} is defined as the set of all tuple indices from $r$ that have identical values for $X$, or $c(t) = \{i | t_i[X] = t[X] \}$. The PLI for $X$ is all such sets, excluding singleton clusters:
\begin{equation*}
    \bar{\pi}(X) = \{c(t) | t \in r \land |c(t)| > 1 \}
\end{equation*}

The size of the resultant index is denoted as $||\bar{\pi}(X)|| = \sum_{c \in \bar{\pi}(X)} |c|$.

Hence, the calculation of the error metric $e$ is as follows: tuple pairs that agree on $X$ and disagree on $A$ (for the candidate $X \to A$) are considered \ul{violating pairs}, which need to be counted. However, instead of counting them directly, PYRO employs a more efficient method. For each cluster $\bar{\pi}(X)$, the number of tuple pairs that also agree on $A$ is calculated, and this result is then subtracted from the total number of tuple pairs in the cluster. This is achieved through $v_A$, a vector in which information about the content of the cluster is recorded in one-hot-encoding format. Summing up the errors for each cluster yields the final error. The pseudocode for this algorithm is presented in Listing~\ref{algo:AFD_PYRO}.

\begin{algorithm}
\begin{algorithmic}
\REQUIRE{Set of tuples $\bar{\pi}(X)$, values of attribute $A$ as $v_A$}
\ENSURE{Metric $e$ for AFD}
\STATE {$e \gets 0$\;}

\FOR{each cluster $c \in \bar{\pi}(X)$}
    \STATE {$counter \gets$ dictionary with default value 0\;}
    \FOR{each item $i \in c$}
        \FOR{$v_A[i] \neq 0$}
            \STATE{$counter[v_A[i]] \gets counter[v_A[i]] + 1$\;}
        \ENDFOR
    \ENDFOR
    \STATE {$e \gets e + |c|^2 - |c| - \sum_{A \in counter} counter[A]^2 - counter[A]$\;}
\ENDFOR

\RETURN{$e$}
\end{algorithmic}

\caption{Calculation of $e$ for AFD using PYRO}
\label{algo:AFD_PYRO}
\end{algorithm}

\textbf{2. Approximate unique column combinations.} \ul{Approximate Unique Column Combinations} (AUCCs) represent another type of relaxed dependency that can be discovered using the PYRO algorithm. 

Let $r$ be a relation with schema $R$ and attribute sets $X,Y \subseteq R$. According to~\cite{10.14778/3192965.3192968}, $X$ is a \ul{Unique Column Combination} (UCC) if, for all tuple pairs $t_1,t_2 \in r$, from $t_1[X] \neq t_2[X]$ it follows that $t_1[Y] = t_2[Y]$.

The error metric for AUCC is defined as follows:
\begin{equation*}
\begin{split}
 e(X \to &A, r) = \\ 
& =\frac{|\{(t_1, t_2) \in r^2 | t_1[X] \neq t_2[X] \land t_1[A] = t_2[A]\}|}{|r|^2 - |r|}. 
\end{split}
\end{equation*}

For Approximate UCC, unlike AFD, the error calculation for $\bar{\pi}(X)$ is trivial. This happens because all tuple pairs within each cluster are the violating tuples themselves.

\begin{algorithm}
\begin{algorithmic}

\REQUIRE{Set of tuples $\bar{\pi}(X)$, total number of tuples $|r|$}
\ENSURE{Metric $e$ for AUCC}

\STATE{$e \gets \sum_{c \in \bar{\pi}(X)} \frac{|c|^2 - |c|}{|r|^2 - |r|}$\;}

\RETURN{$e$}
\end{algorithmic}

\caption{Calculation of $e$ for AUCC using PYRO}
\label{algo:AUCC_PYRO}
\end{algorithm}

Approximate Unique Column Combinations are utilized in tasks such as data cleaning, database normalization, and query optimization.

\textbf{3. Denial Constraints.} \ul{Denial constraint} (DC) is a type of an integrity constraint used in databases to ensure data quality. DC describes conditions that must not occur within the database. For example, it might state that two rows in a table cannot have certain value combinations. If an insertion or an update of a row violates a DC, the operation is generally aborted.

\ul{Approximate denial constraints} in databases are a form of constraint that permits a degree of flexibility or exceptions. Unlike exact denial constraints that rigorously prohibit certain data value combinations, approximate denial constraints allow for a limited number of violations.

In this case, the $g_1$ metric is utilized for calculating the error measure~\cite{pena2019discovery}.

DCs and their approximate variants are essential for upholding data consistency and reliability within a database, as they avert the introduction of invalid or conflicting information.

\subsection{$g_3$ metric}
Let $G_3$ represent the number of tuples for the dependency $X \to Y$ within the relation $r$ that must be removed to establish an exact dependency. Formally:
\begin{equation*}
G_3(X \to Y, r) = |r| - max \{|s| : s \subset r, s \models X \to Y \}
\end{equation*}
\begin{equation*}
g_3(X \to Y, r) = G_3(X \to Y, r) / |r|
\end{equation*}

The $g_3$ metric is acknowledged as an industry standard and is applied in the context of various approximate dependencies: Approximate Functional Dependencies, Approximate Inclusion Dependencies, Probabilistic Functional Dependencies.

Its calculation is as follows. For the example presented in table~\ref{tbl:ExampleCalculation}:
$$g_3(X \to Y, r) = \frac{5 - 4}{5},$$
this is because it is sufficient to remove one tuple for the ``exact'' dependency to be satisfied. This example illustrates the practical utility of the $g_3$ in assessing the degree of violation of a dependency within a dataset. Now, let us consider various relaxed dependencies that are
based on this metric.

\textbf{1. Approximate functional dependencies.} Despite the fact that modern AFD discovery papers utilize the $g_1$ metric, the initial paper proposing the AFD concept employed $g_3$. This paper also proposed the TANE algorithm~\cite{tane}, designed for mining exact functional dependencies which can also be modified for mining approximate functional dependencies. The metric was defined as follows:
\begin{equation*}
e(X \to A) = min \{ \frac{|s|}{|r|} : s \subset r \ and \ X \to A \ holds \ in \ r \setminus s\}
\end{equation*}
Another algorithm~\cite{caruccio2020mining} utilizing the $g_3$ metric is called $DiM\varepsilon$. This highly-optimized algorithm employs a level-wise approach in candidate generation, starting with singleton sets at level zero. Additionally, the authors claim that the algorithm can be adapted for use with other metrics and even different types of dependencies.

Functional and approximate functional dependencies are instrumental in database normalization, data cleaning, and also aid analysts in uncovering hidden trends within data. Their implementation and optimization in algorithms like TANE and $DiM\varepsilon$ highlight their significance in managing and analyzing large data sets efficiently.

\textbf{2. Approximate inclusion dependencies.} An \ul{Inclusion Dependency}~\cite{10.14778/3192965.3192968} (IND) over a schema $R$ is a statement of the form $R_i[X] \subseteq R_j[Y]$, $R_i, R_j \in R$, $X \subseteq R_i$, $Y \subseteq R_j$. The size (or arity) of such a dependency is denoted as $i = R[X] \subseteq R[Y]$, where $|i| = |X| = |Y|$. Inclusion dependencies of size one are commonly referred to as unary inclusion dependencies. 

An inclusion dependency is \ul{satisfied} if all values from the left side are present in the right side. To assess the degree of approximation, a variant of the $g_3$ metric, denoted as $g_3'$, is used. This version is adapted for inclusion dependencies and conveys essentially the same meaning.

Despite the lack of separate algorithms for detecting \ul{approximate inclusion dependencies}, several algorithms for finding ``exact'' dependencies have been adapted for this task, such as MIND~\cite{marchi2009unary}, Spider~\cite{bauckmann2010efficient}, or S-indd~\cite{10.1007/978-3-319-18120-2_25}.

MIND employs a level-wise approach, where candidates of size $i+1$ are generated from already discovered dependencies of size $i$. In the case of approximate dependencies, during the candidate validation stage, approximate dependencies that meet a user-defined threshold for $g_3'$ are also considered.

The primary application area for both ``exact'' and approximate inclusion dependencies is in the identification of foreign keys in databases~\cite{10.1007/3-540-45876-X_30}. This is crucial for database design, integrity, and normalization processes, facilitating effective data management and interrelation of different data sets within a database system. An example of AIND is presented in Table~\ref{tbl:UserEmailToRegisteredEmail}.

\begin{table}[h!]
\caption{AIND Example: User Email $\to$ Registered Email ($\varepsilon = 0.34$)}
\label{tbl:UserEmailToRegisteredEmail}

\begin{center}
\begin{tabular}{|c | c | c | c|} 
\hline
TID & User Email & ID & Registered Email \\ [0.5ex] 
\hline\hline
T001 & example@email.com & C123 & example@email.com \\ 
\hline
T002 & sample@email.com & C124 & sample@email.com \\ 
\hline
T003 & missing@email.com & C125 & - \\ 
\hline
\end{tabular}
\end{center}
\end{table}
\textbf{3. Graph Entity Dependencies.} A \ul{Graph Entity Dependency} (GED) is a constraint within a property graph $G$, expressed as a pair $\phi = (Q[\bar{u}], X \rightarrow Y)$. It states that for any instance of the pattern in the graph $Q[\bar{u}]$ within $G$, the dependency $X \to Y$ must be upheld. This denotes that if specific conditions defined by $X$ are met within a pattern instance, then other conditions outlined by $Y$ must also be satisfied. The metric $g_3$ is employed in its original form as a measure of approximation for GED~\cite{zhou2023fastageds}.

Graph Entity Dependencies are employed for several key objectives within the realm of graph databases and data management. They ensure data integrity and consistency and aid in the optimization of complex queries.

\textbf{4. Approximate Interval-based Temporal Dependencies.}

\ul{Approximate Interval-based Temporal Functional Dependencies} (AITFDs)~\cite{aibtd, 6940375} are a type of constraint in temporal databases. They extend the concept of functional dependencies to consider the temporal aspect of data, specifically focusing on time intervals. They use $g_3$ as a metric of approximation as follows.

Let $X$ and $Y$ be sets of atemporal attributes of a temporal relation schema $R = R(U, B, E)$, $~$ an Allen's Interval relation and $\epsilon$ a real number $0 \leq \epsilon \leq 1$. An instance $r$ of $R$ \ul{satisfies} an ITFD $X \to~ Y$ with approximation $\epsilon$ if there exists a subset $r' \subseteq r$ for which $r \setminus r' \models X \to~  Y$ and $|r'| \leq \epsilon \cdot |r|$.

AITFDs are used for maintaining data integrity in temporal databases by ensuring that relationships among data attributes adhere to specified patterns over time. They are particularly useful for analyzing historical data, identifying trends, and predicting future values by understanding the temporal dynamics of data relationships.

\textbf{Wrap-up.} Concluding this section, we can state that, to the best of our knowledge, there were no studies where comparison between pFDs and AFDs was performed. The reasons for this are the following: 
\begin{enumerate}
    \item Both notions were developed long before the era of data profiling began.
    \item Each notion was developed by a different research group and for a particular task.
    \item The notions were assessed by its applicability to this particular task only, or no comparisons were performed at all.
\end{enumerate}

Currently, data profiling is gaining traction, and it is imperative to catalogue all available tools. Thus, it is essential to compare pFDs and AFDs  with each other.

\section{Algorithms and Implementation}\label{sec:algo}
This paper considers an implementation of pFDTane algorithm designed to discover minimal non-trivial probabilistic functional dependencies. 

The new algorithm is based on TANE~\cite{tane}, which is a graph-traversing algorithm in which a graph~--- called lattice~--- is comprised of vertices representing all possible sets of attributes and edges connecting nodes of a form X and XA, where X~--- set of vertices, and A~--- another attribute. This way every edge represents a functional dependency $X \xrightarrow{}$~$ A$. The algorithm consecutively checks for the existence of functional dependencies between neighboring levels of lattice, excluding vertices whenever possible.

\begin{figure*}[htp]\centering
    \centering
    
    \includegraphics [width=.99\textwidth]{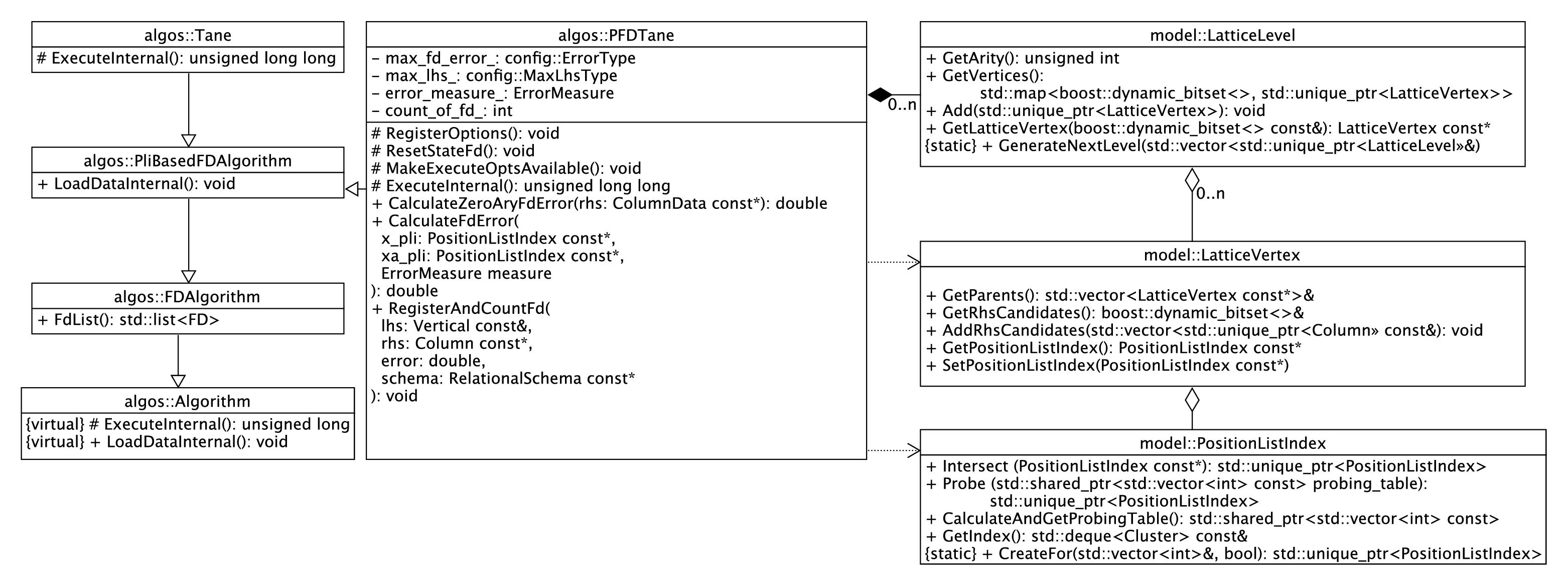}

    \caption{UML class diagram}
    \label{fig:uml}
\end{figure*}

\textbf{Integration.} In Desbordante, FDs discovery algorithms are implemented by inheriting FDAlgorithm or its subclasses and overriding ExecuteInternal method. Tane and PFDTane shown on the diagram in Figure~\ref{fig:uml} inherit PliBasedAlgorithm, in which relation loading method is additionally overridden.  PositionListIndex (PLI) is a useful data structure comprised of stripped partitions~\cite{tane}. This means that the structure contains a set of equivalence classes, built with respect to the equality of attribute values. Stripped means that classes containing a single attribute are dropped to reduce memory consumption. In Desbordante, this set is represented by a double-ended queue~--- namely, std::deque. More specifically, inheritance and related classes are shown on Figure~\ref{fig:uml}.

\begin{algorithm}[H]
\caption{Calculation of PerValue~\cite{wang} metric}
\label{algo:PerValue}
\begin{algorithmic}
    
 \REQUIRE Relation $R$, attributes X and A
 \ENSURE  Metric PerValue for $X\xrightarrow{}A$
    \STATE {$c \gets t_1(X); |\pi(X)| \gets 1 ; count(c) \gets 0$}
    
    \STATE {$c' \gets t_1(X, A); count(c') \gets 0;  maxCount(c) \gets 0$}
    
    \STATE {$sum \gets 0$}
    \FOR{each $ t \in R$}
  \IF{$t(X) == c$} 
    \STATE {$count(c) \gets count(c) + 1$\;}
    \IF{$ t(X, A) == c'$} 
        \STATE {$count(c') \gets count(c') + 1$\;}
      \ELSE 
      
      \IF{$ maxCount(c) < count(c')$} 
        \STATE {$maxCount(c) \gets count(c')$\;}
        \ENDIF

        \STATE {$c' \gets t(X, A); count(c') \gets 0$\;}

      \ENDIF
  \ELSE 
        \STATE {$sum \gets sum + maxCount(c)/count(c)$\;}
        \STATE {$c \gets t(X); |\pi(X)| \gets |\pi(X)| + 1$\;}
        \STATE {$count(c) \gets 0; maxCount(c) \gets 0$}
  \ENDIF
  \ENDFOR
\RETURN{$sum/|\pi(X)|$}
\end{algorithmic}

\end{algorithm}

The class PFDTane uses LatticeLevel and LatticeVertex data structures, which contain the level and vertex information respectively. PFDTane generates lattice levels and handle its life time, so there an aggregation dependency with LatticeLevel  is shown. Meanwhile ExecuteInternal method uses LatticeVertex and PLI, LatticeLevel consists of instances of LatticeVertex and each PLI instance corresponds to LatticeVertex, which is shown as composition on the diagram.

\textbf{Candidate Validation.} Error measurement functions used for candidate validation is the essentially only part which had to be changed in order to adapt the existing TANE implementation for pFD discovery. The functions implements the algorithms presented in listings~\ref{algo:PerValue} and~\ref{algo:PerTuple}. Implemented functions for non-zero FDs take PLI of LHS attributes of dependency and PLI of a union of LHS attributes and RHS attribute as arguments. Sorting performed in the first lines of code in Listings~\ref{algo:PerValue} and~\ref{algo:PerTuple} is done on the latter argument, i.e. union of PLIs. Thus, it is then possible to iterate over PLI clusters, calculating probability in linear time. Because of using PLI, the algorithm does not iterate over single value clusters, which has positive impact on the algorithms run time.

\begin{algorithm}[t]

\caption{Calculation of PerTuple metric}
\label{algo:PerTuple}
\begin{algorithmic}

 \REQUIRE Relation $R$, attributes X and A
 \ENSURE  Metric PerTuple for $X\xrightarrow{}A$

 \STATE {$c \gets t_1(X); count(c) \gets 0$}
\STATE {$c' \gets t_1(X, A); count(c') \gets 0; maxCount(c) \gets 0$}
\STATE {$sum \gets 0 $}

\FOR{each $ t \in R$} 
  \IF{$t(X) == c$} 
    \STATE{$count(c) \gets count(c) + 1$\;}
    \IF{$ t(X, A) == c'$} 
        \STATE {$count(c') \gets count(c') + 1$\;}
      \ELSE 
      
      \IF{$ maxCount(c) < count(c')$} 
        \STATE {$maxCount(c) \gets count(c')$\;}
      \ENDIF
        \STATE {$c' \gets t(X, A); count(c') \gets 0$\;}
      \ENDIF
  \ELSE 
              \STATE {$sum \gets sum + maxCount(c)$\;}
              \STATE {$c \gets t(X); |\pi(X)| \gets |\pi(X)| + 1$\;}
              \STATE {$count(c) \gets 0; maxCount(c) \gets 0$}
  \ENDIF
  
\ENDFOR
\RETURN{$sum/|R|$}
\end{algorithmic}

\end{algorithm}

\section{Evaluation and Discussion}\label{sec:experiments}

\subsection{Methodology and Experimental Setup}

\begin{table}
    \centering
    \parbox[t]{.24\textwidth}{
        \centering
        \caption{Example Case 1}
        \label{tbl:PerValuePositive}
        \begin{tabular}{|c | c|} 
            \hline
            X & Y \\ [0.5ex] 
            \hline\hline
            $0$ & 1  \\ 
            \hline
            $0$ & 2  \\ 
            \hline
            $0$ & 3  \\ 
            \hline
            $0$ & 4  \\ 
            \hline
            $0$ & 5  \\ 
            \hline
            ... & ... \\
            \hline
            1 & 1 \\ 
            \hline
            2 & 2  \\ 
            \hline
            3 & 3  \\ 
            \hline
            4 & 4  \\ 
            \hline
            5 & 5  \\ 
            \hline
            6 & 6  \\ [1ex] 
            \hline
        \end{tabular}
    }
    \parbox[t]{.24\textwidth}{
        \centering
        \caption{Example Case 2}
        \label{tbl:PerValueNegative}
        \begin{tabular}{|c | c|} 
            \hline
            X & Y \\ [0.5ex] 
            \hline\hline
            $0$ & 1  \\ 
            \hline
            $0$ & 1  \\ 
            \hline
            $0$ & 1  \\ 
            \hline
            $0$ & 1  \\ 
            \hline
            $0$ & 1  \\ 
            \hline
            1 & 1 \\ 
            \hline
            1 & 2  \\ 
            \hline
            2 & 3  \\ 
            \hline
            2 & 4  \\ 
            \hline
            3 & 5  \\ 
            \hline
            3 & 6  \\ [1ex] 
            \hline
        \end{tabular}
    }
\end{table}

\textbf{Methodology.} In order to answer research questions posed in the introduction, we have decided to perform quantitative and qualitative studies. For the former, we are going to analyze pFDs using examples and conduct an extensive literature review. For the latter, we are going to run a series of experiments, featuring AFD, pFD PerTuple, and pFD PerValue discovery algorithms. All these algorithms were implemented in Desbordante, and, more specifically, we used our TANE implementation. 
It is necessary to mention that, similarly to Metanome in Desbordante, TANE algorithm can cause a larger search space than necessary. However, due to the specifics of implementation, this functionality incurs almost negligible RAM overhead. Turning to performance, we want to stress the fact that this also does not negatively affects our study, since all experiments either compare methods relatively, or they compare algorithm output.

For pFD PerValue and pFD PerTuple algorithms, each dataset have been run with error thresholds ranging from 0 to 1 (inclusively) with an increment of 0.025. For TANE, datasets have been run with error thresholds located in captions of Tables~\ref{table:exp3a_pv}--\ref{table:exp3b_pt}. This subset was selected due to the fact that error values close to zero are of more value to user and are expected to be used much more frequently.

A total of 10 iterations per error threshold value had been run, after which the average query execution time and maximum memory usage were calculated with confidence interval of 95\%. Due to large confidence intervals for the run time of measures\_v2.csv, an additional set of 20 iterations have been performed for that specific dataset in order to get a more accurate data.

\textbf{Datasets.} To perform experimental evaluation, we used datasets presented in Table~\ref{table:datasets}. Links to the datasets are available in the GitHub repository~\cite{datasets_sources}. To perform comprehensive evaluation, we tried to select a collection of datasets with different properties. In this table we list the number of rows and attributes, file size and file source, as well as the number of minimal non-trivial FDs, AFDs, pFDs (both PerTuple and PerValue). The AFDs and pFDs were calculated with error threshold set to $0.01$.

We have divided these datasets into two groups, which we present separately in two distinct figures. The reason for this is the dataset size difference, which will make them poorly readable if we put them in the same figure.

\textbf{Experimental setup.} Experiments were performed using the following hardware and software configuration. Hardware: AMD® Ryzen 5 7600X CPU @ 5.453GHz (6 cores), 32GB RAM. Software: Ubuntu 22.04 LTS, Kernel 6.5.0-15-generic (64-bit).

\subsection{RQ1: Are pFDs of interest to the end-user? How much of a difference there is between pFDs and AFDs, and what kind of FD violations are pFDs more tolerant to? Does this definition allow for discovery of dependencies that could not be discovered by the AFD definition? And vice versa: how much AFDs are lost by it?}

Let us start with the qualitative comparison of AFDs and pFDs. 

\textbf{Observation 1.} First, lets consider a simplified dataset $R$ presented in Table~\ref{tbl:PerValuePositive} and $X\xrightarrow{}Y$ dependency.

pFD with PerValue metric is not affected by the frequency of X. Indeed, consider $|V_{0}| \xrightarrow{} \infty$. In this case $P_{PerValue}(X \xrightarrow{}$~$Y, R)$ tends to $\frac{6}{7}$ and its respective error $1 - P_{PerValue}(X \xrightarrow{}$~$Y, R) $ tends to $ \frac{1}{7}$. At the same time, $g_3(X \xrightarrow{} Y, R)$ and $e(X \xrightarrow{} Y, R)$ tends to $1$.

Thus, this metric can account for ``faulty'' LHS, if there are not too much of them. It allows having a lot of violating records if they correspond to relatively few distinct LHS values. For example, such ``local'' error may arise if a single sensor of an overall healthy set started to report faulty data.

\textbf{Observation 2.} Now, consider data presented in Table~\ref{tbl:PerValueNegative} and the same dependency.

The dependency is less likely to hold with the PerValue metric:  $P_{PerValue}(X \xrightarrow{}$~$ Y, R) = \frac{5}{8} = 0.625$, and $1 - P_{PerValue}(X \xrightarrow{}$~$ Y, R) = 0.375$. At the same time $g_3(X \xrightarrow{}$~$Y, R) = \frac{3}{11} = 0.27$, and $e(X \xrightarrow{}$~$ Y, R) = \frac{3}{55} = 0.05$.

Thus, pFD's PerValue will report larger error when there are a lot of individual LHS where dependency does not hold. It will ignore the positive contribution of ``0'', regardless of their number.

For a data scientist who explores data, such behavior may be undesirable and lead to valuable facts being missed. Suppose that there are one million of those ``0'' in this table, and the other six entries stay the same. This table will result in PerValue error of $0.375$, which is rather large and therefore the pattern described by this pFD can be ignored. However, a more probable interpretation is the following: one million records are correct (since there is one million of them) and these six values are anomalies which should be deleted. At the same time, such interpretation can be located using AFDs.

\textbf{FD guessing problem.} pFDs were extensively used for problems where the goal was to ``guess'' FDs~\cite{wang, berti2018genuineFDs, Hazar2020disser} from low-quality data. In these studies true FDs (gold standard) were known beforehand and had to be discovered using pFDs. Authors who originally proposed the pFD concept have performed experiments~\cite{wang} which have shown that PerTuple tends to yield better results than PerValue in finding correct dependencies in cases where data is of a lower quality (due to noise).

However, their subsequent experiments~\cite{wang_report} with an improved version of TANE that uses transitivity rule have had PerValue outperforming PerTuple in the majority of cases. The authors measured recall, precision, and F-measure using a gold-standard collection.

Recently, the PerValue metric demonstrated~\cite{berti2018genuineFDs, Hazar2020disser} better results for the problem of FD discovery in datasets containing missing values.

\textbf{AFDs vs pFDs, quantatively.} The above-mentioned studies have not considered AFDs and their difference from pFDs. In our qualitative study we have demonstrated cases where pFDs can be of use and where they are inferior to AFDs. Now, let us turn to quantative part, which aims to answer the rest of the RQ1: ``Does this definition allow for discovery of dependencies that could not be discovered by the AFD definition? And vice versa: how much AFDs are lost by it?''.

Table~\ref{table:pfd_vs_afd_discovery} contains the results of a search for three different dependency types in the monkeypox.csv dataset: AFDs, pFDs with PerValue, and pFDs with PerTuple. The table shows that AFD fails to find some pFDs when run with certain error thresholds, despite the dataset containing a comparable number of minimal non-trivial AFDs.

Though the minimal sets indeed differ, it doesn't immediately imply that the complete sets of pFDs and AFDs do. For example, for a fixed threshold, you may have found the following minimal dependencies: $pfd_1: XZ\rightarrow A,\quad pfd_2: XY\rightarrow A  $ and $afd_1: X \rightarrow A$. But  $afd_1$ infers all other AFDs that have $X$ in LHS. That implies $pfd_1, pfd_2$ are in set of \textit{all} AFDs. In order to highlight the essential difference of pFDs, we have also included in the table the number of minimal pFDs that are neither in the set of minimal AFDs nor inferable from it.

Concluding this RQ, we can say that pFDs have their own strengths, and that they are different from AFDs. Specifically, having fixed error threshold, pFDs are not a mere subset of AFDs, nor are AFDs a subset of pFDs in general. Finally, existing studies have demonstrated that pFDs have found applications for the FD guessing problem. However, in those studies comparison with AFDs was not performed, and it is outside of scope of this paper.

\begin{table*}[ht]
	\begin{center}

	\caption{Minimal non-trivial pFDs and AFDs found in monkeypox.csv}
	\label{table:pfd_vs_afd_discovery}
		\begin{tabular}{|c|c|c|c|c|c|c|c|c|c|c|c|}
			\hline
			 & \multicolumn{3}{|c|}{Total} & \multicolumn{2}{|c|}{$|pFDs\setminus AFDs|$} & \multicolumn{2}{|c|}{Non-inferable pFDs} & \multicolumn{2}{|c|}{$|AFDs\setminus pFDs|$} & \multicolumn{2}{|c|}{$|pFDs \cap AFDs|$} \\
            \hline

   			Error & AFD & pFD PerValue & pFD PerTuple & PerValue & PerTuple & PerValue & PerTuple & PerValue & PerTuple & PerValue & PerTuple  \\
            \hline

            0.01 & 126 & 142 & 134 & 133 & 124 & 3 & 1 & 117 & 116 & 9 & 10 \\
            \hline
            0.05 & 73 & 69 & 71 & 62 & 64 & 2 & 2 & 66 & 66 & 7 & 7 \\
            \hline
            0.1 & 55 & 81 & 64 & 72 & 55 & 2 & 0 & 46 & 46 & 9 & 9 \\
            \hline
            0.2 & 69 & 168 & 70 & 153 & 60 & 29 & 0 & 54 & 59 & 15 & 10 \\
            \hline
            0.3 & 63 & 70 & 51 & 61 & 42 & 30 & 2 & 54 & 54 & 9 & 9 \\
            \hline
  \end{tabular}
	\end{center}
\end{table*}

\subsection{RQ2: How computationally expensive is the candidate validation procedure of pFD discovery algorithm compared to the AFD?}

To compare the run time and memory consumption of validation functions of pFDTane and AFDTane, the corresponding algorithms had been run with error set to $0$. This setting guarantees that all algorithms traverse the same part of the lattice and they all are on the level playing field. The results presented in Table~\ref{table:exp1} showcase the fact that both PerValue and PerTuple prove to work slower than the validation with $g_1$. On the other hand, PerValue and PerTuple do not demonstrate a significant difference in either run time or memory consumption.

We can also note that almost all datasets have had less memory consumed by pFDTane when compared to AFDTane. The exception~--- which was the SEA.csv dataset~--- used approximately the same amount of memory.

\begin{table*}[ht]
	\begin{center}
 
	\caption{Exact FDs discovery time and memory}
	\label{table:exp1}
		\begin{tabular}{|c|c|c|c|c|c|c|}
			\hline
			\multirow{2}{*}{Datasets} & \multicolumn{3}{c|}{Time (s)} & \multicolumn{3}{c|}{Memory (MB)} \\
			\cline{2-7}
			   & pFDTane per\_value & pFDTane per\_tuple & AFDTane & pFDTane per\_value & pFDTane per\_tuple & AFDTane \\
			\hline
			BKB\_WaterQualityData\_2020084 & 2.013 & 2.018 & 0.935 & 216 & 216 & 260 \\
			\hline
			EpicVitals & 8.609 & 8.583 & 4.266 & 807 & 807 & 1490 \\
			\hline
			jena\_climate\_2009\_2016 & 12.711 & 12.720 & 6.205 & 530 & 530 & 1490 \\
			\hline
			measures\_v2 & 16.245 & 16.334 & 15.278 & 1758 & 1758 & 1785 \\
			\hline
			nuclear\_explosions & 2.198 & 2.203 & 0.795 & 189 & 189 & 1490 \\
			\hline
			parking\_citations & 25.908 & 25.953 & 7.360 & 1381 & 1381 & 1491 \\
			\hline
			SEA & 1.963 & 1.956 & 1.193 & 279 & 279 & 270 \\
			\hline
			games & 3.207 & 3.222 & 1.492 & 399 & 399 & 1490 \\
			\hline
		\end{tabular}

  	\end{center}
\end{table*}

\subsection{RQ3: How does maximum error threshold affect run time and memory consumption of the pFD discovery algorithm?}

\begin{table*}[htp]
    \centering

 	\caption{Datasets used for experiments}
	\label{table:datasets}
        \setlength\tabcolsep{5pt} 
		\begin{tabular}{|c|c|c|c|c|c|c|c|}
		\hline
        Dataset & Rows & Attributes & Size & Source & pFD PT count & pFD PV count & AFD count \\
		\hline
        EpicVitals.csv & 1246303 & 7 & 33MB & EPF & 10 & 13 & 21 \\
        \hline
        BKB\_WaterQualityData\_2020084.csv & 2370 & 17 & 180KB & U.S. FWS & 3389 & 3712 & 901  \\
        \hline
        games.csv & 20058 & 16 & 7.67MB & kaggle  & 2264 & 1810 & 266 \\
        \hline
        jena\_climate\_2009\_2016.csv & 420550 & 15 & 43.16MB & kaggle  & 3003 & 3148 & 210 \\
        \hline
        measures\_v2.csv & 1330816 & 13 & 300.06MB & kaggle  & 642 & 573 & 144 \\
        \hline
        nuclear\_explosions.csv & 
        2046 & 16 & 220KB & tidytudesday repository  & 2795 & 3619 & 1459 \\
        \hline
        parking\_citations.csv & 
        95433 & 13 & 10MB & norfolk opendata  & 224 & 269 & 565 \\

        \hline 
        SEA.csv & 1000000 & 4 & 33MB &  openml.com & 3 & 3 & 9 \\

        \hline
		monkeypox.csv & 5875 & 14 & 516KB & who.int & 134&142&126   \\
  \hline
  \end{tabular}
    
\end{table*}

\begin{figure}
\centering
\subfloat[EpicVitals\label{fig:e2a_epicvitals}]{\includegraphics[height=.2\textheight]{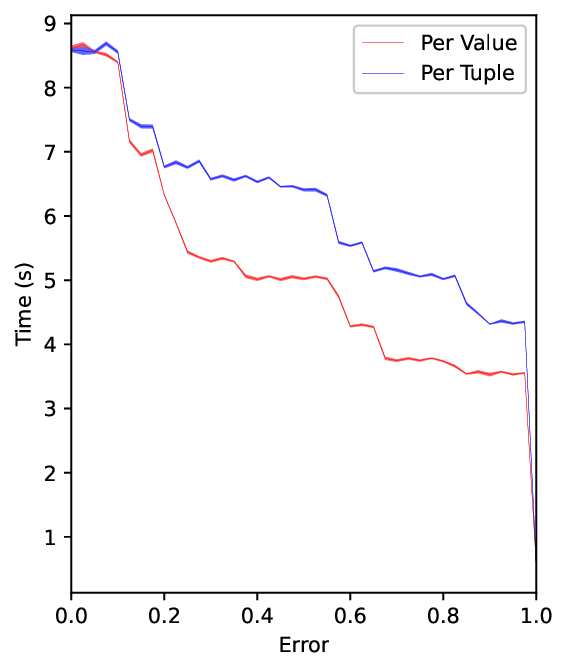}}
\hfill
\subfloat[jena\_climate\_2009\_2016\label{fig:e2a_jena}]{\includegraphics[height=.2\textheight]{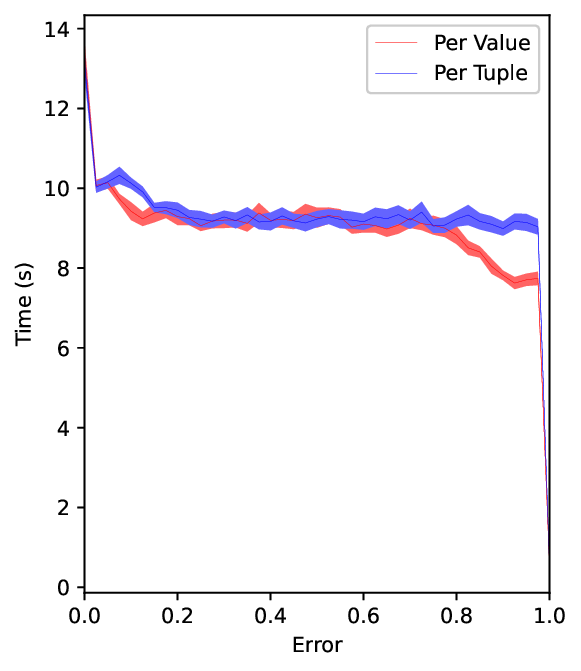}}
\caption{pFDTane running times}
\label{fig:decreasing_vs_constant_time}
\end{figure}

The pFDTane algorithm have been run with various error thresholds ranging from 0 to 1 on eight different datasets depicted in Table~\ref{table:datasets}. Figure~\ref{fig:e2a_epicvitals} and Figure~\ref{fig:e2a_jena} show two different patterns of behaviours of pFDTane. When it's supplied with the jena\_climate\_2009\_2016.csv dataset, run time does not demonstrate a noteworthy difference past the 0.25 error threshold. Contrary to that, when run on the EpicVitals.csv dataset, the algorithm gets progressively faster as the error threshold increases. In our experiments six out of eight datasets behaved similarly to EpicVitals.

The number of steps in Tane is determined by the number of vertices in the lattice. However, the algorithm discards some vertices during its execution due to the nature of the task of searching for the minimal functional dependencies. Thus, the observed trend could be explained by the difference between 0.2 and 0.8 error threshold not generating any new dependencies, which would subsequently lead to an inability to discard additional vertices in the lattice.

\begin{figure*}
\centering
\subfloat[First Batch\label{fig:time_1}]{\includegraphics[height=.6\textheight]{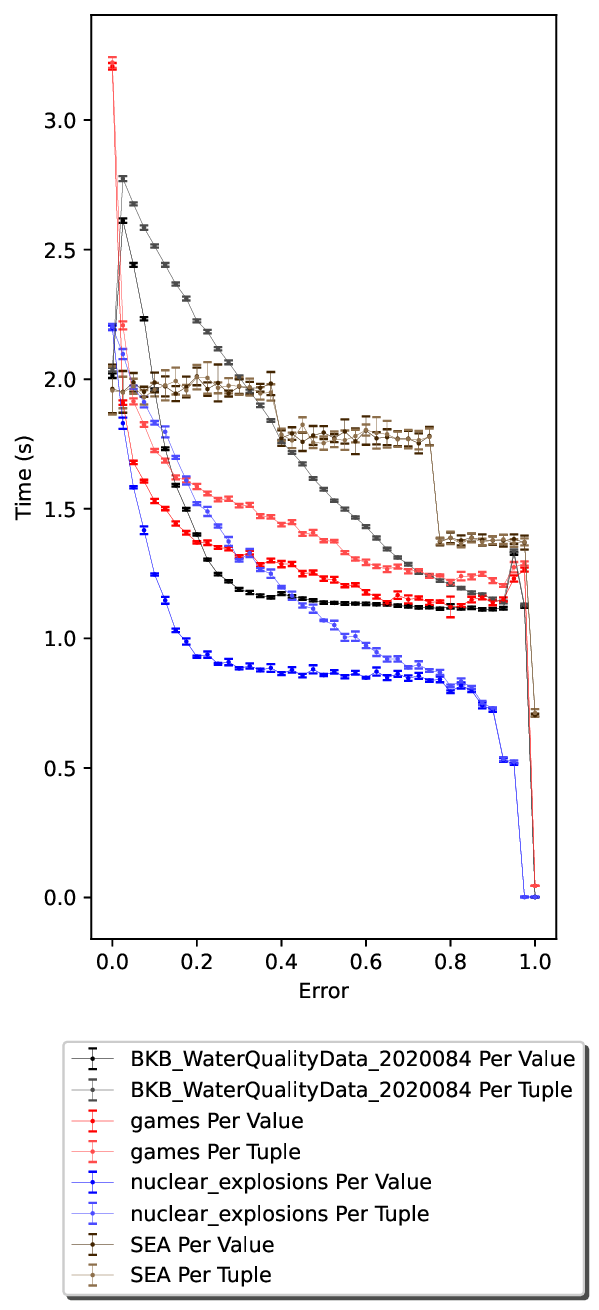}}
\subfloat[Second Batch\label{fig:time_2}]{\includegraphics[height=.6\textheight]{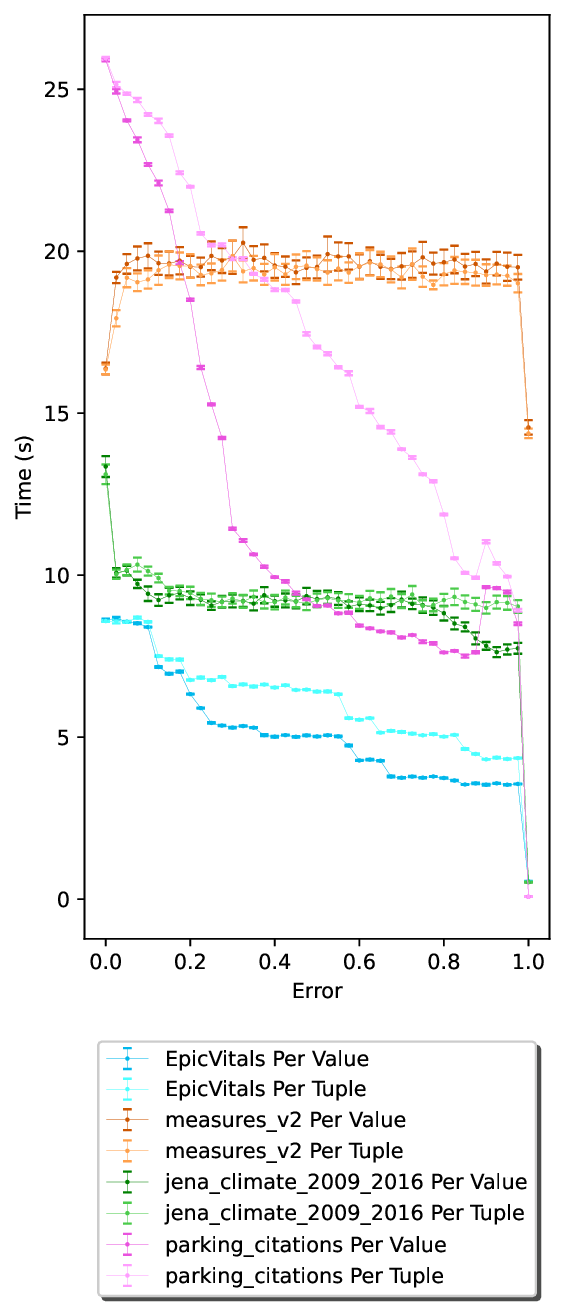}}
\caption{Running time by error threshold}
\label{fig:all_time} 
\end{figure*}

Finally, as could be observed from Figure~\ref{fig:all_time}, PerValue yields better results in terms of run time on every dataset but SEA.csv when compared to pFDs with PerTuple.

\begin{figure}
\centering
    \subfloat[First Batch\label{fig:memory_1}]{\includegraphics[height=.35\textheight]{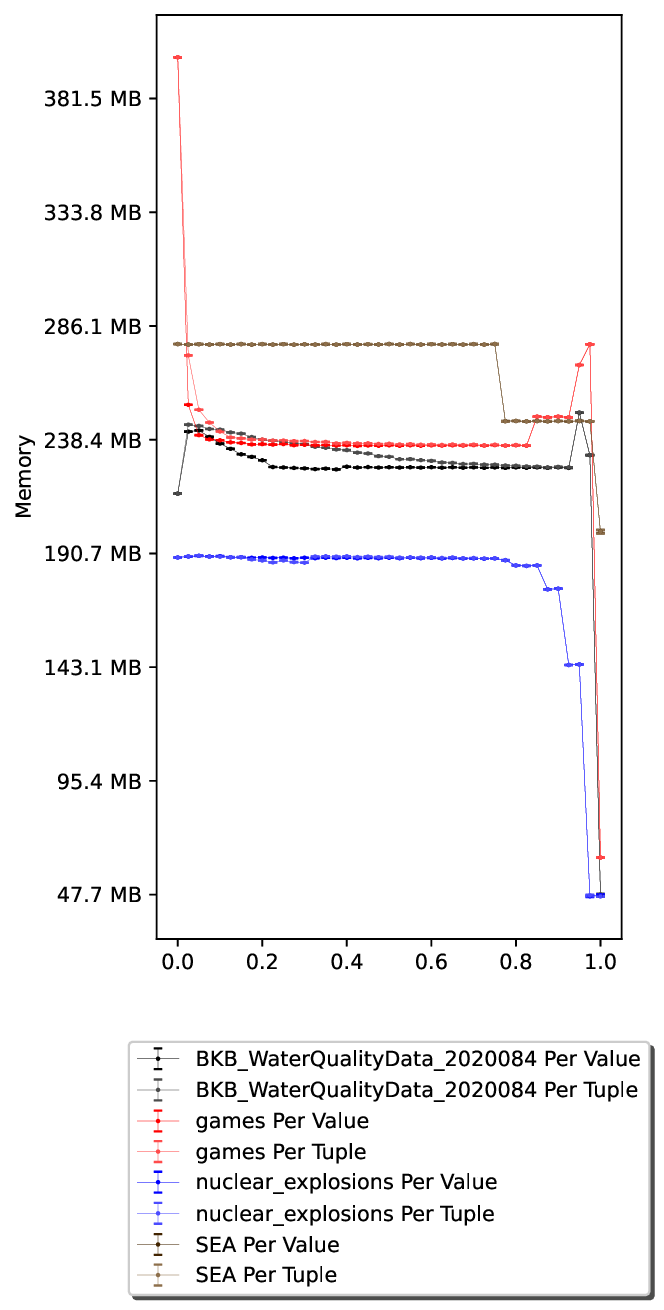}}
\hfill
\subfloat[Second Batch\label{fig:memory_2}]{\includegraphics[height=.35\textheight]{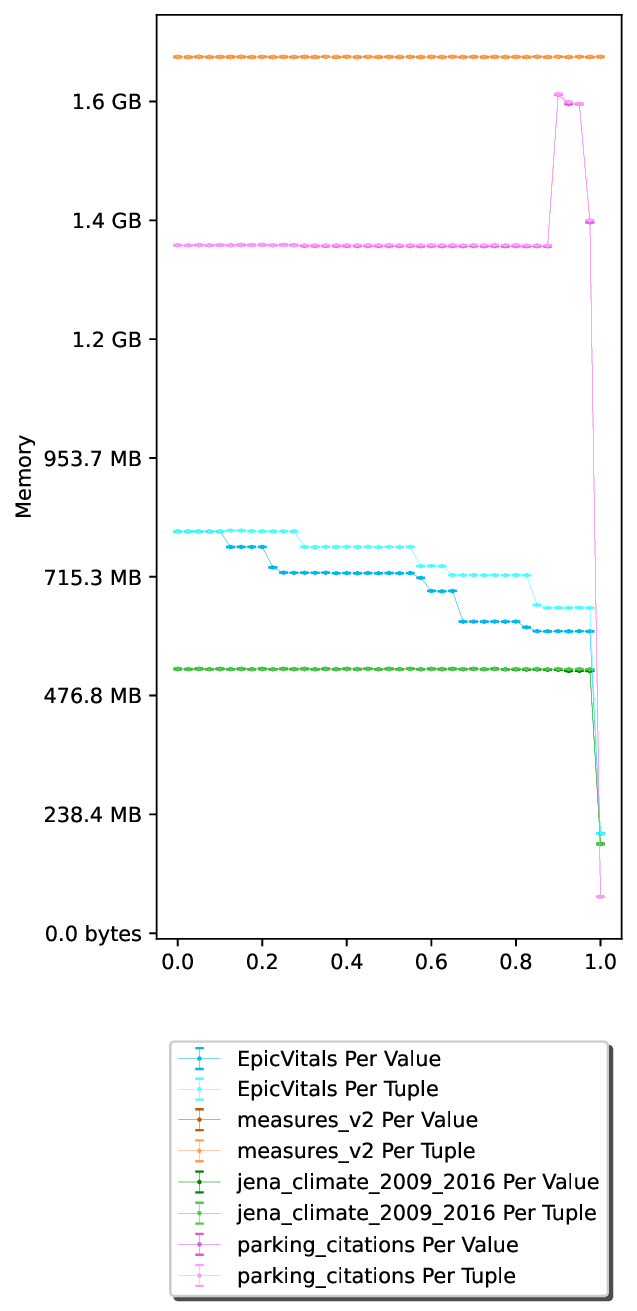}}
\caption{Maximum memory consumption by error threshold}
\label{fig:all_memory}
\end{figure}

\subsection{RQ4: What is the run time and memory expenses of pFD discovery, compared to AFD?}

To compare pFDTane and AFDTane algorithms in the probabilistic and approximate dependency discovery tasks, the algorithms' implementations have been tested with different error thresholds. The results are presented in Table~\ref{table:exp3a_pv}, Table~\ref{table:exp3a_pt} for the run time and in Table~\ref{table:exp3b_pv} and Table~\ref{table:exp3b_pt} for the memory consumption respectively. Each cell contains the ratio of pFDTane to AFDTane respective measurements. For the ease of understanding we have plotted the maximum amount memory consumed graph in Figure~\ref{fig:all_memory}.

Almost all of the datasets depict AFDTane as a faster discovery algorithm when compared to pFDTane, with an exception of measures\_v2.csv. The results suggest a decrease in performance difference with the error threshold exceeding 0.1.

Memory consumption has been observed to be lower for pFDTane on all datasets but SEA.csv. In contrast to the run time metric, the memory consumption seems to be equal for pFDTane and AFDTane for each of error threshold values. 

\begin{table*}[htp]
	\centering
 
    \caption{Ratio of running time of AFDTane and pFDTane PerValue algorithms}
    \label{table:exp3a_pv}
    
    \begin{tabular}{|c|c|c|c|c|c|c|c|c|c|c|}
        \hline
        \diagbox{Dataset}{Error threshold} & 0.025 & 0.05 & 0.075 & 0.1 & 0.15 & 0.2 & 0.25 & 0.3 & 0.4 & 0.5 \\
        \hline
        BKB\_WaterQualityData\_2020084 & 2.529 & 2.199 & 2.010 & 1.763 & 1.432 & 1.267 & 1.128 & 1.076 & 1.050 & 1.027 \\
        \hline
        EpicVitals & 2.671 & 2.637 & 2.623 & 2.589 & 2.145 & 1.952 & 1.679 & 1.636 & 1.548 & 1.547 \\
        \hline
        jena\_climate\_2009\_2016 & 1.455 & 1.475 & 1.470 & 1.359 & 1.387 & 1.316 & 1.308 & 1.322 & 1.320 & 1.360 \\
        \hline
        measures\_v2 & 0.962 & 0.972 & 0.951 & 0.921 & 0.946 & 0.953 & 0.992 & 0.985 & 0.949 & 0.915 \\
        \hline
        nuclear\_explosions & 2.160 & 1.862 & 1.668 & 1.466 & 1.220 & 1.101 & 1.070 & 1.048 & 1.024 & 1.015 \\
        \hline
        parking\_citations & 3.358 & 3.261 & 3.177 & 3.079 & 2.882 & 2.513 & 2.073 & 1.550 & 1.351 & 1.231 \\
        \hline
        SEA & 1.745 & 1.815 & 1.767 & 1.805 & 1.752 & 1.821 & 1.794 & 1.773 & 1.615 & 1.674 \\
        \hline
        games & 1.785 & 1.587 & 1.517 & 1.446 & 1.361 & 1.299 & 1.277 & 1.217 & 1.212 & 1.162 \\
        \hline
    \end{tabular}
\end{table*}

\begin{table*}[htp]
	\centering
 
    \caption{Ratio of running time of AFDTane and pFDTane PerTuple algorithms}    \label{table:exp3a_pt}
    
    \begin{tabular}{|c|c|c|c|c|c|c|c|c|c|c|}
        \hline
        \diagbox{Dataset}{Error threshold} & 0.025 & 0.05 & 0.075 & 0.1 & 0.15 & 0.2 & 0.25 & 0.3 & 0.4 & 0.5 \\
        \hline
        BKB\_WaterQualityData\_2020084 & 2.685 & 2.415 & 2.320 & 2.262 & 2.127 & 2.007 & 1.908 & 1.804 & 1.578 & 1.417 \\
        \hline
        EpicVitals & 2.642 & 2.636 & 2.677 & 2.638 & 2.283 & 2.087 & 2.086 & 2.031 & 2.017 & 1.974 \\
        \hline
        jena\_climate\_2009\_2016 & 1.465 & 1.479 & 1.514 & 1.487 & 1.410 & 1.343 & 1.326 & 1.396 & 1.316 & 1.368 \\
        \hline
        measures\_v2 & 0.894 & 0.952 & 0.904 & 0.892 & 0.928 & 0.948 & 0.961 & 0.965 & 0.943 & 0.914 \\
        \hline
        nuclear\_explosions & 2.475 & 2.316 & 2.251 & 2.153 & 2.010 & 1.802 & 1.701 & 1.542 & 1.419 & 1.264 \\
        \hline
        parking\_citations & 3.386 & 3.373 & 3.344 & 3.289 & 3.197 & 2.987 & 2.740 & 2.681 & 2.557 & 2.318 \\
        \hline
        SEA & 1.742 & 1.800 & 1.747 & 1.780 & 1.796 & 1.819 & 1.777 & 1.776 & 1.626 & 1.636 \\
        \hline
        games & 2.064 & 1.810 & 1.723 & 1.630 & 1.529 & 1.503 & 1.451 & 1.399 & 1.356 & 1.300 \\
        \hline
    \end{tabular}
\end{table*}

\begin{table*}[htp]
	\centering

    \caption{Ratio of consumed memory of AFDTane and pFDTane PerValue algorithms} 	\label{table:exp3b_pv}
    
    \begin{tabular}{|c|c|c|c|c|c|c|c|c|c|c|}
        \hline
        \diagbox{Dataset}{Error threshold} & 0.025 & 0.05 & 0.075 & 0.1 & 0.15 & 0.2 & 0.25 & 0.3 & 0.4 & 0.5 \\
        \hline
        BKB\_WaterQualityData\_2020084 & 0.931 & 0.932 & 0.922 & 0.911 & 0.894 & 0.885 & 0.873 & 0.871 & 0.874 & 0.873 \\
        \hline
        EpicVitals & 0.542 & 0.542 & 0.542 & 0.542 & 0.521 & 0.521 & 0.486 & 0.486 & 0.486 & 0.486 \\
        \hline
        jena\_climate\_2009\_2016 & 0.356 & 0.356 & 0.356 & 0.356 & 0.356 & 0.356 & 0.356 & 0.356 & 0.356 & 0.356 \\
        \hline
        measures\_v2 & 0.985 & 0.985 & 0.985 & 0.985 & 0.985 & 0.985 & 0.985 & 0.985 & 0.985 & 0.985 \\
        \hline
        nuclear\_explosions & 0.127 & 0.128 & 0.127 & 0.127 & 0.127 & 0.127 & 0.127 & 0.127 & 0.127 & 0.127 \\
        \hline
        parking\_citations & 0.926 & 0.926 & 0.926 & 0.926 & 0.926 & 0.926 & 0.926 & 0.925 & 0.925 & 0.925 \\
        \hline
        SEA & 1.029 & 1.030 & 1.029 & 1.030 & 1.030 & 1.030 & 1.030 & 1.030 & 1.030 & 1.046 \\
        \hline
        games & 0.170 & 0.162 & 0.160 & 0.160 & 0.159 & 0.159 & 0.159 & 0.159 & 0.159 & 0.159 \\
        \hline
    \end{tabular}
\end{table*}

\begin{table*}[htp]
	\centering

    \caption{Ratio of consumed memory of AFDTane and pFDTane PerTuple algorithms} 	\label{table:exp3b_pt}
    
   \begin{tabular}{|c|c|c|c|c|c|c|c|c|c|c|}
        \hline
        \diagbox{Dataset}{Error threshold} & 0.025 & 0.05 & 0.075 & 0.1 & 0.15 & 0.2 & 0.25 & 0.3 & 0.4 & 0.5 \\
        \hline
        BKB\_WaterQualityData\_2020084 & 0.942 & 0.940 & 0.935 & 0.934 & 0.927 & 0.918 & 0.915 & 0.911 & 0.901 & 0.890 \\
        \hline
        EpicVitals & 0.542 & 0.542 & 0.542 & 0.542 & 0.543 & 0.542 & 0.542 & 0.521 & 0.521 & 0.521 \\
        \hline
        jena\_climate\_2009\_2016 & 0.356 & 0.356 & 0.356 & 0.356 & 0.356 & 0.356 & 0.356 & 0.356 & 0.356 & 0.356 \\
        \hline
        measures\_v2 & 0.985 & 0.985 & 0.985 & 0.985 & 0.985 & 0.985 & 0.985 & 0.985 & 0.985 & 0.985 \\
        \hline
        nuclear\_explosions & 0.127 & 0.128 & 0.127 & 0.127 & 0.127 & 0.126 & 0.126 & 0.126 & 0.127 & 0.127 \\
        \hline
        parking\_citations & 0.926 & 0.926 & 0.926 & 0.926 & 0.926 & 0.926 & 0.926 & 0.926 & 0.926 & 0.926 \\
        \hline
        SEA & 1.029 & 1.030 & 1.029 & 1.030 & 1.030 & 1.030 & 1.030 & 1.030 & 1.030 & 1.046 \\
        \hline
        games & 0.184 & 0.169 & 0.165 & 0.163 & 0.161 & 0.160 & 0.160 & 0.160 & 0.159 & 0.159 \\
        \hline
    \end{tabular}
\end{table*}

\section{Conclusion}\label{sec:concl}

We started with qualitative analysis of pFDs, as well as showing cases in which they have the edge over AFDs and vice versa. Essentially, we demonstrated that data interpretation and data context leave room for both of them, since neither can substitute the other one. Ultimately, it's up to a data scientist to decide what to consider a violation of an exact FD, and the two concepts allow its user to target different cases. Experiments have also shown that pFD is capable of discovering some dependencies that AFD fails to find. The results featured in Table~\ref{table:pfd_vs_afd_discovery} demonstrate that fact by counting discovered dependencies with the same error threshold for both algorithms.

As for performance, the discovery of both pFD types is almost always considerably slower than AFD. However, the memory consumption shows the opposite trend, with pFD using less memory compared to AFD. The difference between run time of pFD and AFD decreases as the error threshold increases, though useful information is primarily mined with a low error threshold. Experiments have also shown similar run time and memory consumption for both pFD PerTuple and pFD PerValue.

Overall, we have introduced pFD discovery functionality into Desbordante for both PerValue and PerTuple metrics, as we had shown that their utility depends on data interpretation and context. At the same time, while building a science-intensive data profiler it is an imperative to expand the catalogue of available tools. Therefore, we hope that this primitive will become another useful tool which will allow our users to uncover knowledge hidden in data. Source code of the implementation is available in the GitHub repository (PR 300)~\cite{desbordante_repo}.

\bibliographystyle{IEEEtran}

\bibliography{FRUCTexample-short}

\end{document}